\documentclass[sigconf]{acmart}
\makeatletter
\AtBeginDocument{%
  \fancyhf{}
  
}
\makeatother

\renewcommand\footnotetextcopyrightpermission[1]{}

\usepackage{subcaption} 

\usepackage{amsmath}
\usepackage{algorithmic}
\usepackage{graphicx}
\usepackage{textcomp}
\usepackage{array}
\usepackage{verbatim}
\usepackage{booktabs}
\usepackage{multirow}
\usepackage{makecell}
\usepackage{caption}
\usepackage[table]{xcolor}
\usepackage{textcomp}
\usepackage{stfloats}

\AtBeginDocument{%
  }
\acmConference{}{}{}
\acmBooktitle{}
\acmDOI{}
\acmISBN{}
\begin{document}

\title{
SliceMoE: Bit-Sliced Expert Caching under \\
Miss-Rate Constraints for Efficient MoE Inference
}


\author{Yuseon Choi, Sangjin Kim, Jungjun Oh, Gwangtae Park, Byeongcheol Kim, and Hoi-Jun Yoo}
\email{{yuseon.choi, sangjinkim, ojj1245, gwangtaepark, bc27.kim, hjyoo}@kaist.ac.kr}
\affiliation{
  \institution{Korea Advanced Institute of Science and Technology (KAIST), Daejeon, Republic of Korea}
  \city{}
  \country{}
}

\begin{abstract}
MoE models offer efficient scaling through conditional computation, but their large parameter size and expensive expert offloading make on-device deployment challenging. Existing acceleration techniques such as prefetching or expert clustering often increase energy usage or reduce expert diversity. We present SliceMoE, an energy-efficient MoE inference framework for miss-rate–constrained deployment. SliceMoE introduces Dynamic Bit-Sliced Caching (DBSC), which caches experts at slice-level granularity and assigns precision on demand to expand effective expert capacity. To support mixed-precision experts without memory duplication, we propose Calibration-Free Asymmetric Matryoshka Quantization (AMAT), a truncation-based scheme that maintains compatibility between low-bit and high-bit slices. We further introduce Predictive Cache Warmup (PCW) to reduce early-decode cold misses by reshaping cache contents during prefill. Evaluated on DeepSeek-V2-Lite and Qwen1.5-MoE-A2.7B, SliceMoE reduces decode-stage energy consumption by up to 2.37× and 2.85×, respectively, and improves decode latency by up to 1.81× and 1.64×, while preserving near–high-bit accuracy. These results demonstrate that slice-level caching enables efficient on-device MoE deployment.

\end{abstract}

\keywords{
Mixture-of-Experts (MoE) inference, bit-sliced architecture, mixed-precision, and cache-aware routing
}
\maketitle

\section{Introduction}

The recent success of large language models (LLMs) \cite{llama, llama2, llama3} has been driven by growing parameter counts following empirical scaling laws~\cite{scalinglaw}, but such growth is ultimately limited by the proportional increase in computation cost. Mixture-of-Experts (MoE) offers an attractive alternative~\cite{sparsely_gated_moe, mixtral} by replacing a single large feed-forward layer with numerous smaller experts and activating only a small subset for each token. This conditional computation provides significantly higher parameter capacity per unit activation cost, and recent MoE-based architectures \cite{qwen15_moe, deepseek_v2, gpt-oss, llama4} demonstrate that MoE is becoming an inevitable design trend across models ranging from several billion to hundreds of billions of parameters.

However, despite reducing activated parameters and computation, MoE retains a large total parameter capacity—often tens of billions—which exceeds the DRAM budget of resource-constrained on-premises devices. As illustrated in Fig.~1(a), practical deployments therefore rely on expert caching: frequently used experts are stored in DRAM, while the full expert set is placed in lower memory hierarchies such as mobile Flash storage, and cache misses trigger Flash-to-DRAM transfers \cite{cacheprior, ssd_offloading_energy}. Because DRAM-to-XPU access is an order of magnitude faster and over 50× more energy-efficient than Flash accesses, even moderate miss rates lead to prohibitive latency and energy overheads.

\begin{figure}[t]
    \centering

    \begin{subfigure}{\linewidth}
        \centering
        \includegraphics[width=\linewidth]{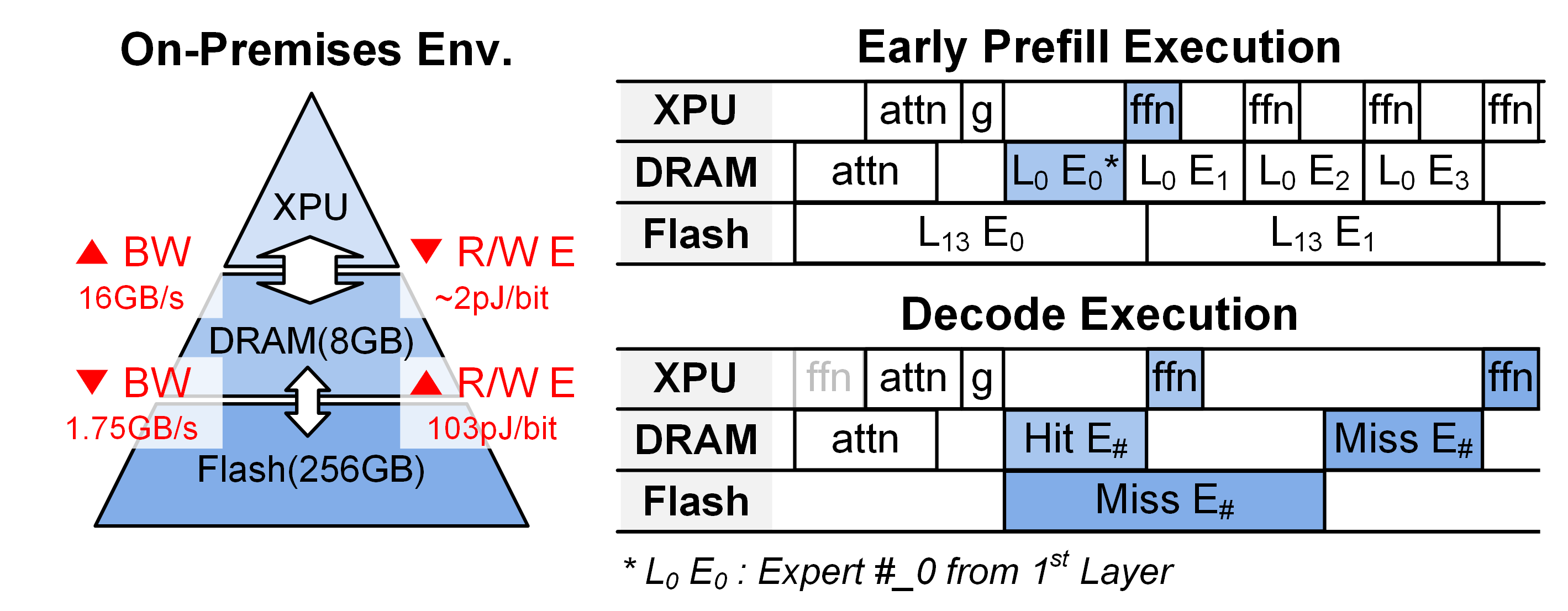}
        \caption{}
        \label{fig_1_a}
    \end{subfigure}

    \begin{subfigure}{\linewidth}
        \centering
        \includegraphics[width=\linewidth]{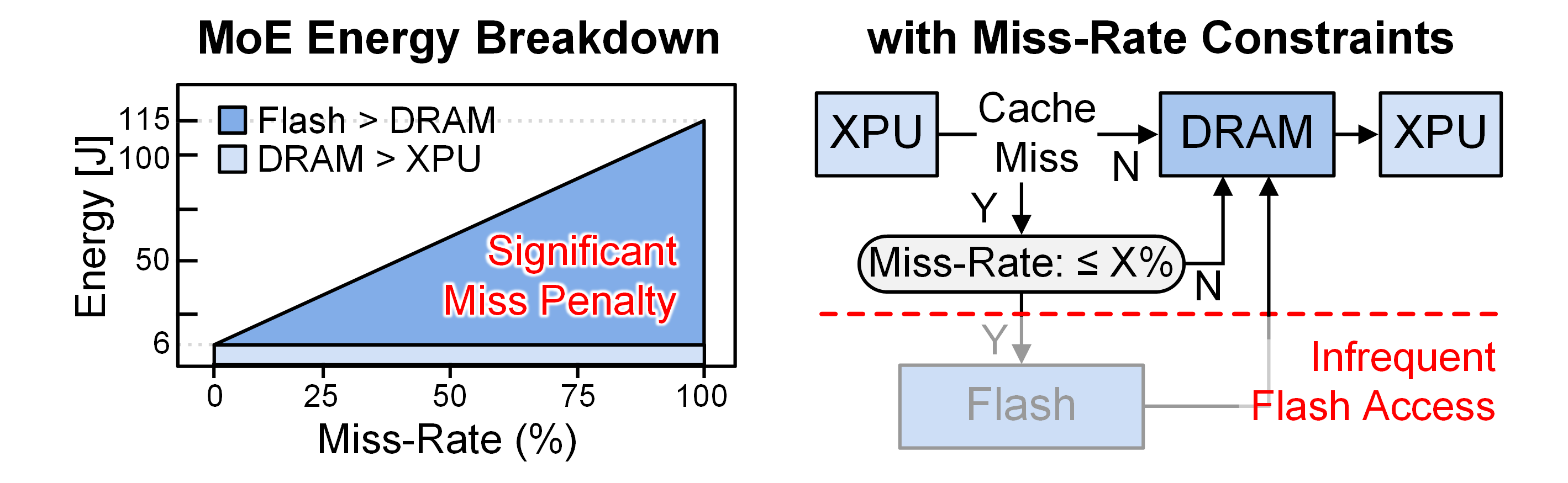}
        \caption{}
        \label{fig_1_b}
    \end{subfigure}

    \caption{(a) On-premises MoE deployment under a single-batch execution scenario. (b) Miss penalty from Flash access and the conceptual execution flow of miss-rate–constrained MoE inference.}
    \label{fig_1}
    \vspace{-0.5cm}
\end{figure}

Algorithmic efforts have attempted to mitigate this bottleneck. Predictive schemes such as prefetching and speculative caching~\cite{pregated_moe, speculative_moe, emoe, promoe, adapmoe} improve locality but become increasingly unreliable in modern MoE. Recent MoE architectures intentionally weaken locality by applying strong router regularization—e.g., load-balancing losses, entropy-boosting terms, and router dropout—to promote expert diversity~\cite{omoe, dive_moe, mask_moe}, which leads to stochastic routing patterns and frequent prefetch failures. Other approaches exploit expert similarity via merging, decomposition, or clustering~\cite{smore, sub_moe, deltamoe, emnlp23_meo, icml25_moe_svd}, but these require modifying model parameters and thus lack generality, limiting applicability to recent high-performance MoE models.

To avoid prediction-based heuristics and parameter modifications, cache-aware routing~\cite{cacheprior, buddymoe, importance_edge_moe} emerges as a robust alternative. Cache-Prior~\cite{cacheprior} improves locality by prioritizing experts already cached in DRAM, reducing miss rates with minimal accuracy degradation. While effective under sufficient cache sizes and moderate miss-rate regions (10–30\%), its accuracy collapses when enforcing the low miss-rate regime (<5\%) required for feasible DRAM–Flash deployment, as shown in Fig.~1(b). Therefore, deploying MoE under realistic mobile memory hierarchies requires a routing and caching solution optimized specifically for extreme low-miss-rate constraints.

To address these challenges, we propose SliceMoE, an energy-efficient MoE inference framework designed around strict miss-rate constraints. Our key contributions are summarized as follows:

\begin{itemize}
    \item \textbf{Dynamic Bit-Sliced Caching:} Slice-level expert caching that supports dynamic precision routing based on gating scores to increase effective expert count and maintain accuracy in a hardware-efficient manner.

    \item \textbf{Calibration-Free Asymmetric Matryoshka Quantization:} A scalable and linear multi-precision scheme that controls zero-points for distribution-adaptive quantization without calibration or memory duplication.
    
    \item \textbf{Predictive Cache Warmup for Cold-Miss Reduction:} A phase-aware warmup strategy that leverages prefill hotness and bit-sliced configuration to reduce early-decode cold misses, improving energy efficiency and reducing latency.
\end{itemize}

\section{Background}

\begin{figure}[t]
    \centering
    \includegraphics[width=\linewidth]{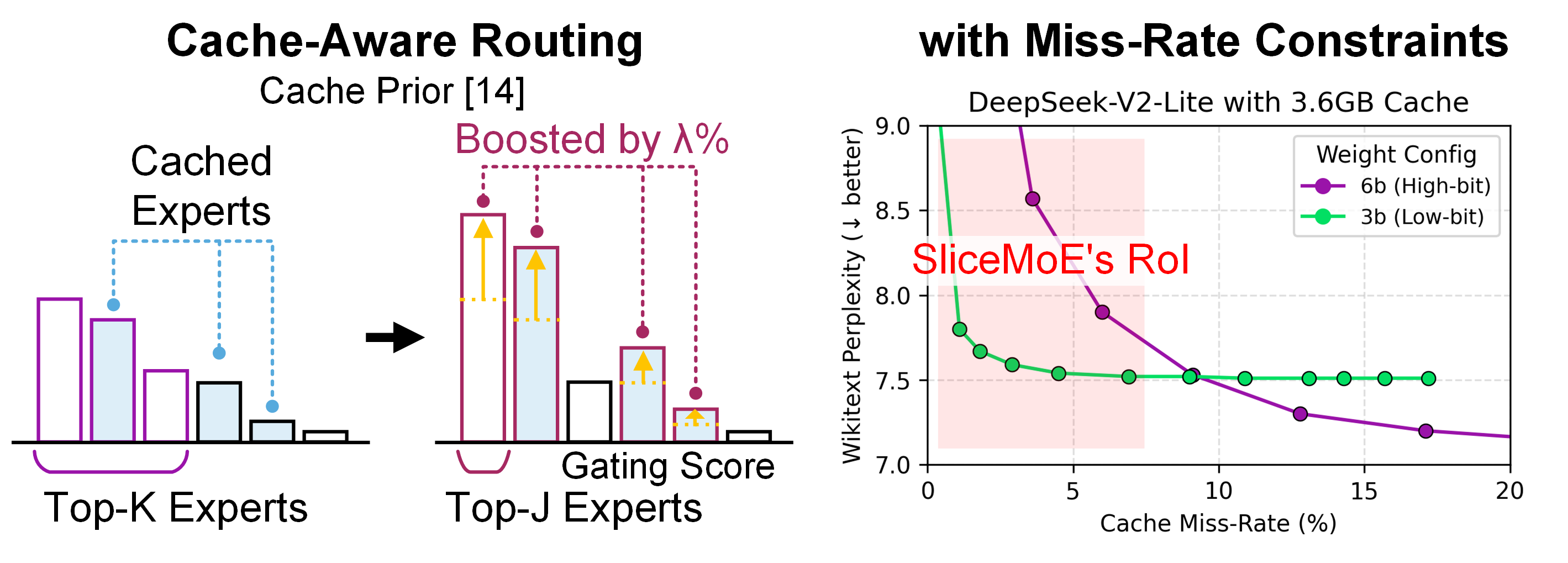}
    \caption{Previous cache-aware routing approach (Cache-Prior) and its limitations within our Region of Interest (RoI) under miss-rate constraints for energy-efficient MoE inference.}
    \label{fig_2}
    \vspace{-0.5cm}
\end{figure}

\begin{figure}[t]
    \centering
    \includegraphics[width=\linewidth]{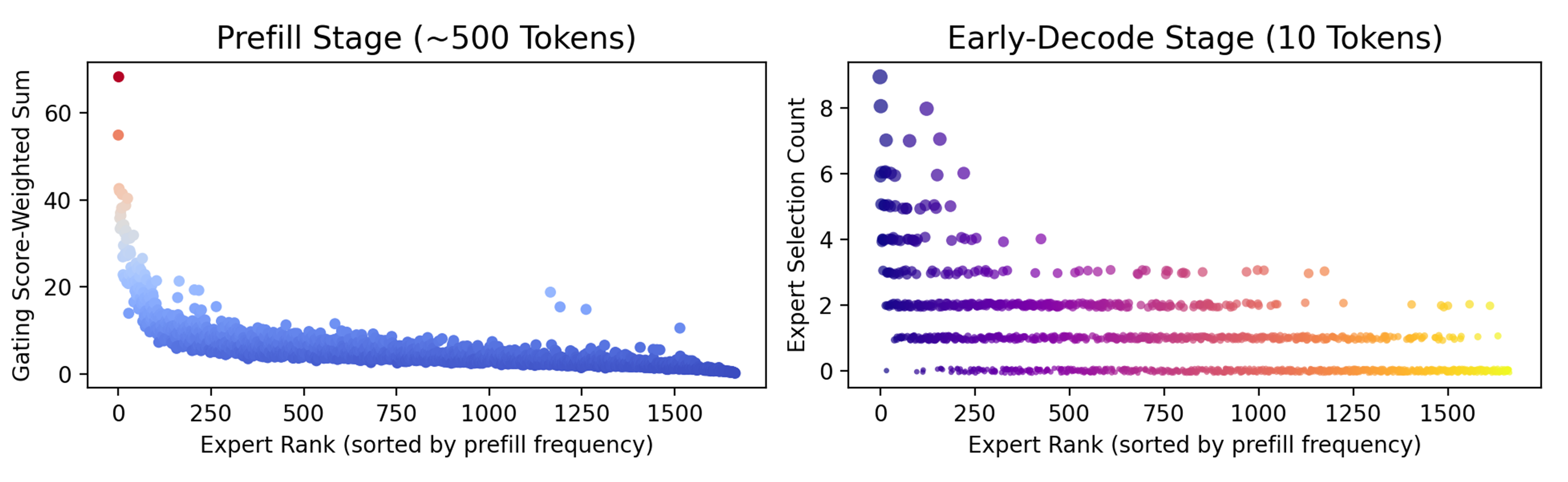}
    \caption{Phase-wise statistics of expert selection frequencies during Prefill and early Decode.}
    \label{fig_2_1}
    \vspace{-0.5cm}
\end{figure}

\subsection{Cache-Aware Routing}

Cache-aware routing~\cite{cacheprior, buddymoe, importance_edge_moe} is built on the observation that experts exhibit partially overlapping coverage across tokens, meaning that certain experts can effectively replace one another during routing. By intentionally biasing routing decisions toward experts already resident in DRAM, these approaches enforce locality and suppress costly Flash accesses.
Cumsum~\cite{cacheprior} selects a candidate expert set based on top gating scores and prioritizes those already cached in DRAM. Cache-Prior~\cite{cacheprior} further formalizes this idea by boosting the gating scores of cached experts, increasing their selection likelihood, as illustrated in Fig. 2(Left). BuddyMoE~\cite{buddymoe} extends this concept by offline-calibrating replacement pairs—experts that are interchangeable for routing—allowing a missed expert to be substituted by a cached one.
These methods consistently outperform simple LRU-based caching, lowering miss rates and reducing Flash accesses. However, when evaluated under realistic energy budgets and strict DRAM–Flash cost asymmetry, their miss rates remain higher than what practical on-device MoE deployment can tolerate.
\subsection{Mixed Precision}

\begin{figure*}[t]
    \centering
    \includegraphics[width=\linewidth]{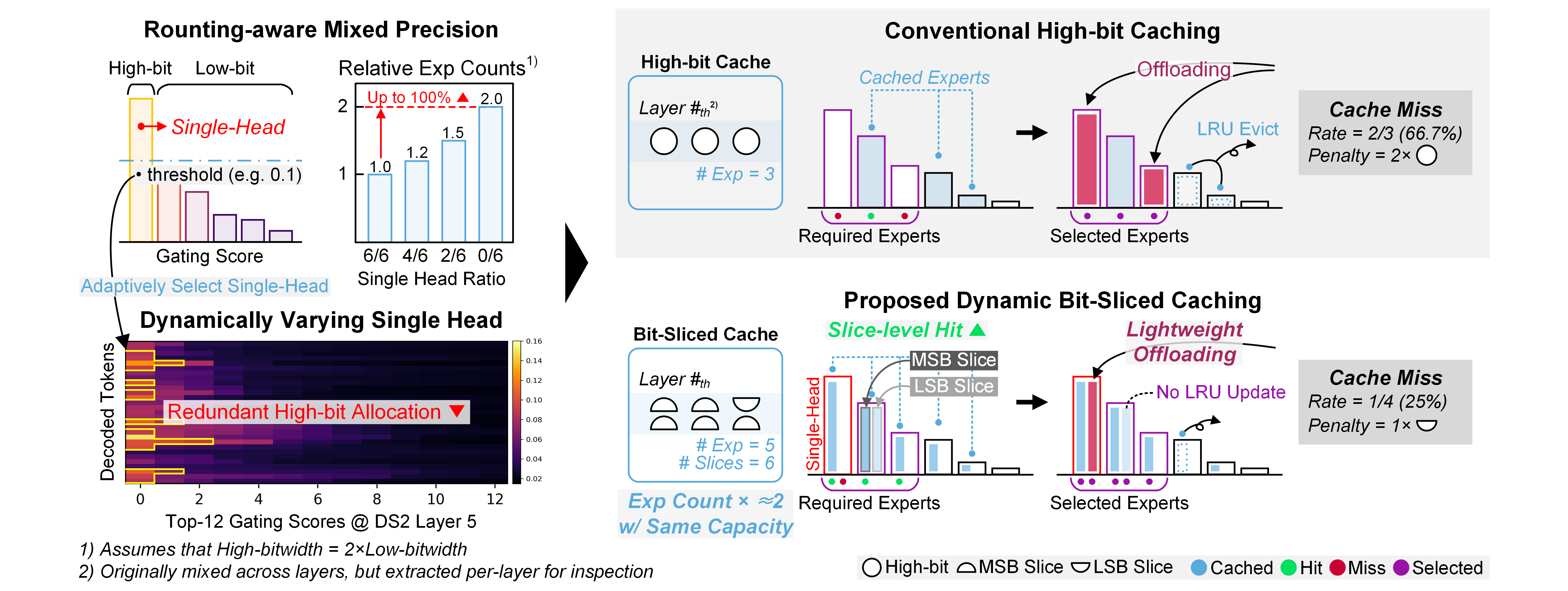}
    \caption{Motivation and overall execution flow of proposed Dynamic Bit-Sliced Caching (DBSC)}
    \label{fig_3}
\end{figure*}

Mixed-precision strategies are also highly effective in MoE systems. Beyond choosing which expert to activate, assigning different bitwidths to different experts improves memory footprint, caching capacity, and energy efficiency.

HOBBIT~\cite{hobbit} adopts a mixed-precision routing policy by allocating precision according to expert importance and adjusting caching/prefetching decisions accordingly. However, this requires storing both high-bit and low-bit copies of weights, leading to memory duplication.

D2MoE~\cite{d2moe} overcomes this limitation by employing non-uniform quantization (NUQ)-based Matryoshka quantization~\cite{anyprecision}, in which lower-bit representations are embedded within the most significant bits of higher-bit weights. This eliminates weight duplication but relies heavily on calibration and NUQ-specific tuning.
\section{Motivation}
MoE inference on-device or on-premises systems operates on a three-tier memory hierarchy—XPU < DRAM (LPDDR4) < UFS 3.1 (Flash). Because Flash accesses are an order of magnitude slower and over tens of times more energy-consuming than DRAM, frequent expert misses quickly dominate latency and energy. Existing caching and prefetching techniques reduce misses but still remain above 10\%, which is impractical for Flash-constrained environments. These limitations reveal two key opportunities.

\textbf{Opportunity 1 — Caching more low-bit experts can outperform caching fewer high-bit experts.}
Miss rate is primarily determined by how many experts fit in DRAM. Since memory capacity scales with bit precision, reducing expert precision allows significantly more experts to be cached. Although low-bit quantization introduces accuracy loss in isolation, the resulting miss-rate reduction often outweighs it. As shown in Fig. 2 (Right), under strict miss-rate constraints (<5\%), caching more low-bit experts achieves higher end-to-end accuracy than caching fewer full-precision ones.

Bit-sliced representations further expand this benefit: experts with low gating scores can be processed using low-bit weights, while highly activated experts can retain higher precision. This selective allocation creates new accuracy–efficiency trade-off points that uniform precision schemes cannot reach.

\textbf{Opportunity 2 — Prefill routing patterns provide strong priors for early decode.}
Prefill and decode exhibit contrasting access patterns. Prefill processes many tokens in parallel and sequentially loads all experts of each layer, while decode caches only a small subset of frequently used experts. This mismatch causes cold misses at the prefill-to-decode transition.

However, as shown in Fig. 3, experts with high gating scores during prefill tend to remain important in early decode. This motivates a prefill-aware cache initialization, where the cache is primed using prefill statistics to avoid early-stage cold misses.

\section{Proposed SliceMoE}

\subsection{Dynamic Bit-Sliced Caching (DBSC)}

Lowering the bit precision of experts increases the number of cacheable experts, thereby improving hit probability under strict miss-rate constraints. However, uniformly reducing precision for all experts imposes an accuracy ceiling. To address this, we employ a mixed-precision strategy where experts use either high-bit or low-bit representations depending on their gating scores. Because gating scores typically follow a steep descending distribution, only a small subset of top-ranked experts significantly influences accuracy, while the majority of non-critical experts can be executed in low-bit precision with minimal impact. This preserves accuracy while effectively expanding cache capacity and reducing offloading penalties.

Furthermore, gating scores often exhibit single-head sharpness~\cite{laser}, meaning the number of truly critical experts varies across tokens. Existing cache-aware routing schemes \cite{cacheprior} assume a fixed top-k set for high-precision allocation, which leads to redundant high-bit usage. As illustrated in Fig. 4(left), token-wise fluctuation (0–2 critical experts) reveals an opportunity to dynamically adjust precision rather than relying on static routing–precision coupling.

To exploit this behavior, we propose Dynamic Bit-Sliced Caching (DBSC), which independently manages MSB and LSB slices to satisfy precision demands at runtime. As shown in Fig. 4(right), critical experts request both MSB and LSB slices, while non-critical experts require only the MSB slice; each slice produces separate hit/miss outcomes. Full precision is reconstructed when both slices are cached, and MSB-only computation is used otherwise. Because critical experts exhibit weaker temporal locality than non-critical ones, the LSB slice shows inherently lower locality than the MSB. We therefore apply heterogeneous cache management: MSB slices follow a standard LRU policy, while LSB slices are assigned the lowest priority and are aggressively evicted after initial access. This slice-level granularity increases the number of effectively cacheable experts, improves slice-level hit rates, and significantly reduces Flash traffic by mitigating miss penalties at the slice level.

\begin{figure}[t]
    \centering
    \includegraphics[width=\linewidth]{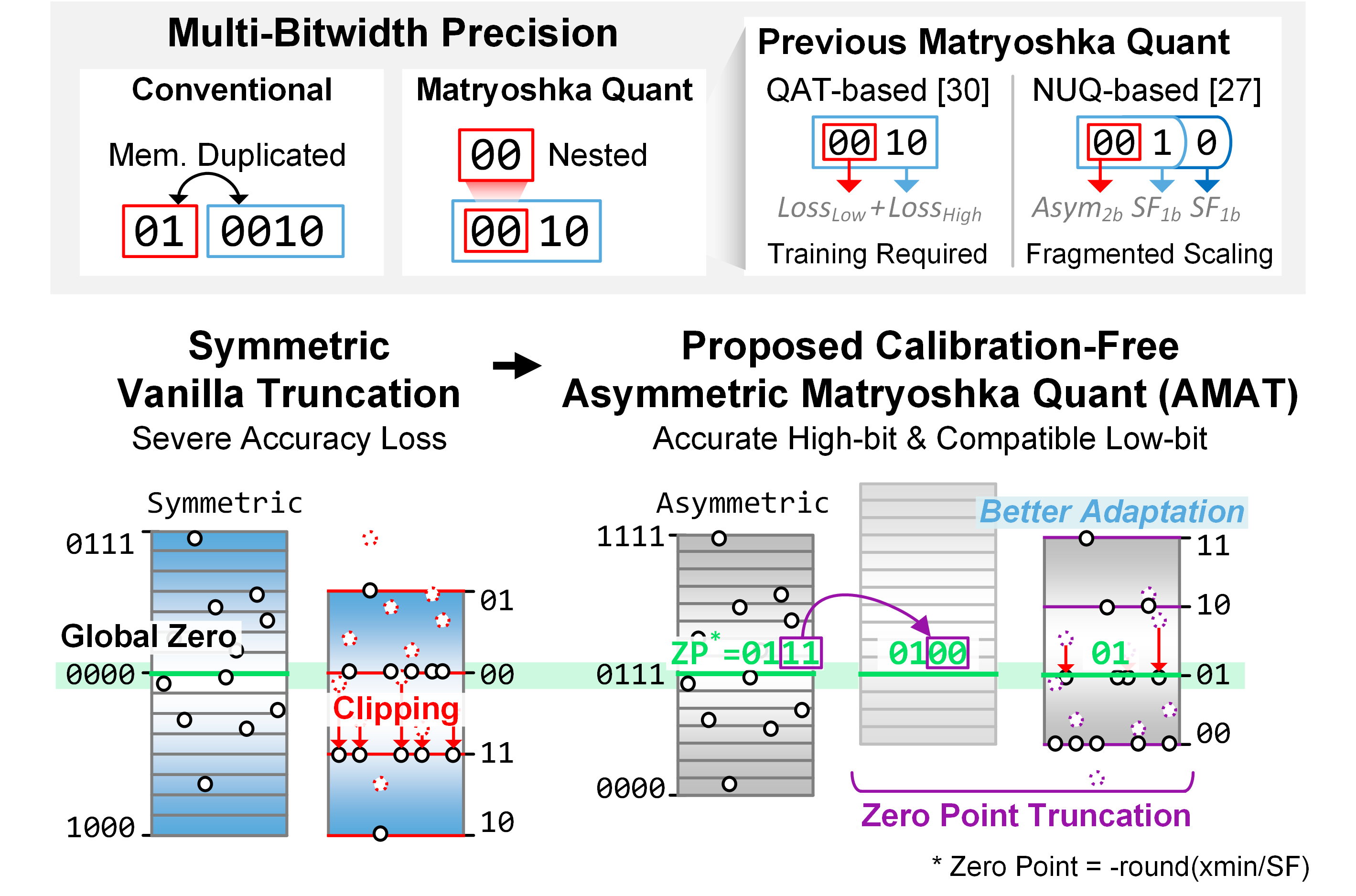}
    \caption{Conventional multi-bitwidth precision methods and the proposed calibration-free Asymmetric Matryoshka Quantization (AMAT).}
    \label{fig_4}
    \vspace{-0.5cm}
\end{figure}

\subsection{Calibration-Free Asymmetric Matryoshka Quantization (AMAT)}

Supporting multi-bitwidth experts typically requires multiple quantized copies of the same parameters, which is impractical in memory-constrained environments. Matryoshka quantization solves this by nesting a low-bit representation inside a high-bit representation, enabling multi-precision inference without duplicating memory. However, existing Matryoshka approaches, such as QAT~\cite{matryoshka} and NUQ~\cite{d2moe, anyprecision}, are impractical due to their complexity and reduced reproducibility. This motivates a simpler, calibration-free linear quantization scheme suitable for deployment.

\begin{table}[]
\centering
\caption{AMAT Accuracy (PPL) on MoE Models}
\vspace{-6pt}
\resizebox{\linewidth}{!}{
\begin{tabular}{cccccccccc}
\toprule
\multirow{2}{*}{\textbf{Models}}
& \multirow{2}{*}{\textbf{Quant$^{1)}$}}
& \multirow{2}{*}{\textbf{Schemes}}
& \multicolumn{2}{c}{\textbf{MAT42$^{2)}$}} 
& \multicolumn{2}{c}{\textbf{MAT63}}
& \multicolumn{2}{c}{\textbf{MAT84}} \\
\cmidrule(lr){4-5} \cmidrule(lr){6-7} \cmidrule(lr){8-9}
& & & \makecell{4b\\[-2pt]} & \makecell{2b\\[-2pt]} 
  & \makecell{6b\\[-2pt]} & \makecell{3b\\[-2pt]}
  & \makecell{8b\\[-2pt]} & \makecell{4b\\[-2pt]} \\
\midrule
\multirow{5}{*}{Deepseek-V2-Lite}
    & \multirow{2}{*}{Sym}
        & Base$^{3)}$  & 7.08 & 19.67 & 7.01 & 7.61 & 7.00 & 7.08 \\
    &     & Trunc$^{4)}$ & 7.08 & 3.6e10 & 7.01 & 3.7e10 & 7.00 & 3.8e10 \\
\cmidrule(lr){2-9}
    & \multirow{3}{*}{Asym}
        & Base  & 7.06 & 9.46 & 7.01 & 7.29 & 7.00 & 7.06 \\
    &     & Trunc & 7.06 & nan & 7.01 & 1.2e9 & 7.00 & 1.2e9 \\
\rowcolor[HTML]{E0F2F5}
    &     & \textbf{AMAT$^{5)}$} & \textbf{7.06} & \textbf{10.18} & \textbf{7.01} 
                          & \textbf{7.56} & \textbf{7.00} & \textbf{7.11} \\
\midrule
\multirow{5}{*}{Qwen1.5-MoE}
    & \multirow{2}{*}{Sym}
        & Base  & 8.18 & 19.75 & 7.97 & 9.12 & 7.97 & 8.18 \\
    &     & Trunc & 8.18 & 1.5e6 & 7.97 & 1.5e6 & 7.97 & 1.5e6 \\
\cmidrule(lr){2-9}
    & \multirow{3}{*}{Asym}
        & Base  & 8.14 & 11.51 & 7.97 & 8.51 & 7.96 & 8.14 \\
    &     & Trunc & 8.14 & 1.8e7 & 7.97 & 1.8e7 & 7.96 & 1.7e7 \\
\rowcolor[HTML]{E0F2F5}
    &     & \textbf{AMAT} & \textbf{8.14} & \textbf{10.85} & \textbf{7.97} 
                   & \textbf{8.61} & \textbf{7.96} & \textbf{8.09} \\
\bottomrule
\end{tabular}
}
\vspace{2pt}
\caption*{\scriptsize
1) Expert-only quantization: activations/weights-FP16; experts-G32 asymmetric INT. \\
2) MAT(h,l) denotes a Matryoshka configuration with high-bit h and low-bit l. \\
3) Base indicates independently quantized bit configurations. \\
4) Trunc refers to standard low-bit truncation. \\
5) AMAT denotes our asymmetric scheme that jointly truncates weights and zero-points.
}
\vspace{-12pt}
\end{table}

Simple truncation provides a path to linear Matryoshka quantization without any calibration. However, a vanilla symmetric truncation is not viable because it introduces substantial clipping error: many values collapse to the truncated boundaries, losing the representability of the original data. This also creates a severely imbalanced quantization range, further biasing the truncated values and degrading low-bit fidelity, as shown in Fig. 5(left).

To address these issues, we introduce Calibration-Free Asymmetric Matryoshka Quantization (AMAT), whose key idea is to truncate the zero-point together with the quantized value, as illustrated in Fig. 5(right). In practice, this means applying the same bit-offset reduction to both terms during precision conversion, 
\[
shift = {b_{\text{high}}}-{b_{\text{low}}},
\qquad
q_{\text{low}}^{\text{trunc}} = \left\lfloor \frac{q_{\text{high}}}{2^{\text{shift}}} \right\rfloor,
\qquad
zp_{\text{low}}^{\text{trunc}} = \left\lfloor \frac{zp_{\text{high}}}{2^{\text{shift}}} \right\rfloor
\]
which re-centers the low-bit range around the asymmetric weight distribution, avoiding the heavy clipping and distortion seen in naive symmetric truncation. Since the resulting low-bit slice behaves much like a properly clipped low-bit quantizer, AMAT preserves meaningful resolution while maintaining an accurate high-bit path fully compatible with bit-sliced caching.

As shown in Table 1, models using low-bit truncation alone suffer severe degradation, whereas AMAT consistently produces low-bit outputs that remain compatible and even outperform the baseline low-bit quantization. Since AMAT only applies zero-point–aware truncation, it requires no calibration or training and reduces memory footprint overhead, resulting in a simple, reproducible, and highly scalable scheme. These properties make AMAT particularly effective for SliceMoE.

\subsection{Predictive Cache Warmup (PCW)}

In single-batch on-premises inference, the cache state differs substantially between the prefill and decode phases. Fig. 6 shows that this discrepancy arises from their different execution mechanisms: prefill operates layer-wise with token-level parallelism, whereas decode proceeds token-by-token with repeated expert loading. As a result, naive initial states—Empty, Last-layer only, or Random retention—perform poorly in early decode, where cold misses dominate both latency and energy, especially under low miss-rate regimes.

To reduce this mismatch, we employ a predictive cache warmup strategy that progressively reshapes the unified cache during prefill. As shown in Fig. 3, experts that are frequently activated during prefill tend to remain important in the early decode phase, providing a clear opportunity for prefill-guided eviction during the phase transition. Slices with consistently low gating scores are discarded first—starting from LSB slices that contribute least to accuracy. After removing these low-sensitivity slices, MSB slices with low prefill access frequency are evicted next. This reflects the observation that frequently accessed MSB slices are far more likely to participate in early decode. Guided by the single-head threshold used in our precision selection, the resulting ratio of experts that retain their MSB slices remains below one on average, ensuring that only a small fraction of experts preserve high-bit precision while the majority are safely reduced to low-bit form.

\begin{figure}[t]
    \centering
    \begin{subfigure}{\linewidth}
        \centering
        \includegraphics[width=\linewidth]{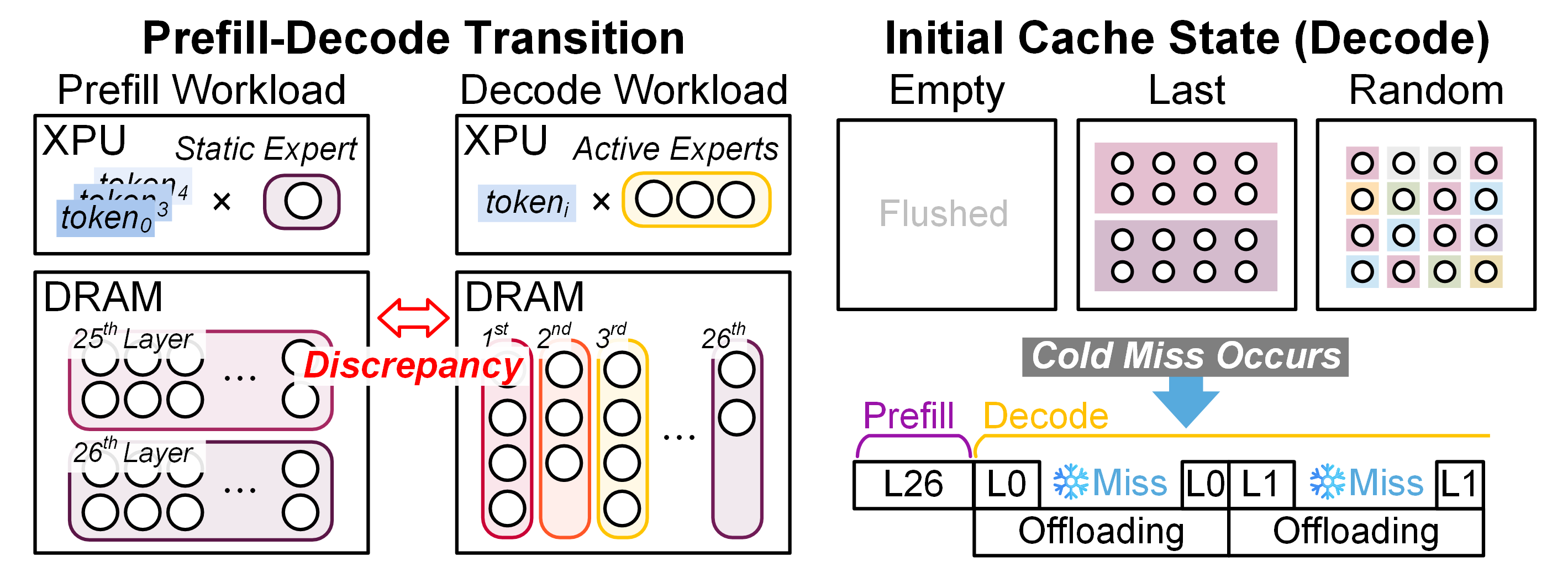}
        \caption{}
        \label{fig_5_a}
    \end{subfigure}
    \begin{subfigure}{\linewidth}
        \centering
        \includegraphics[width=\linewidth]{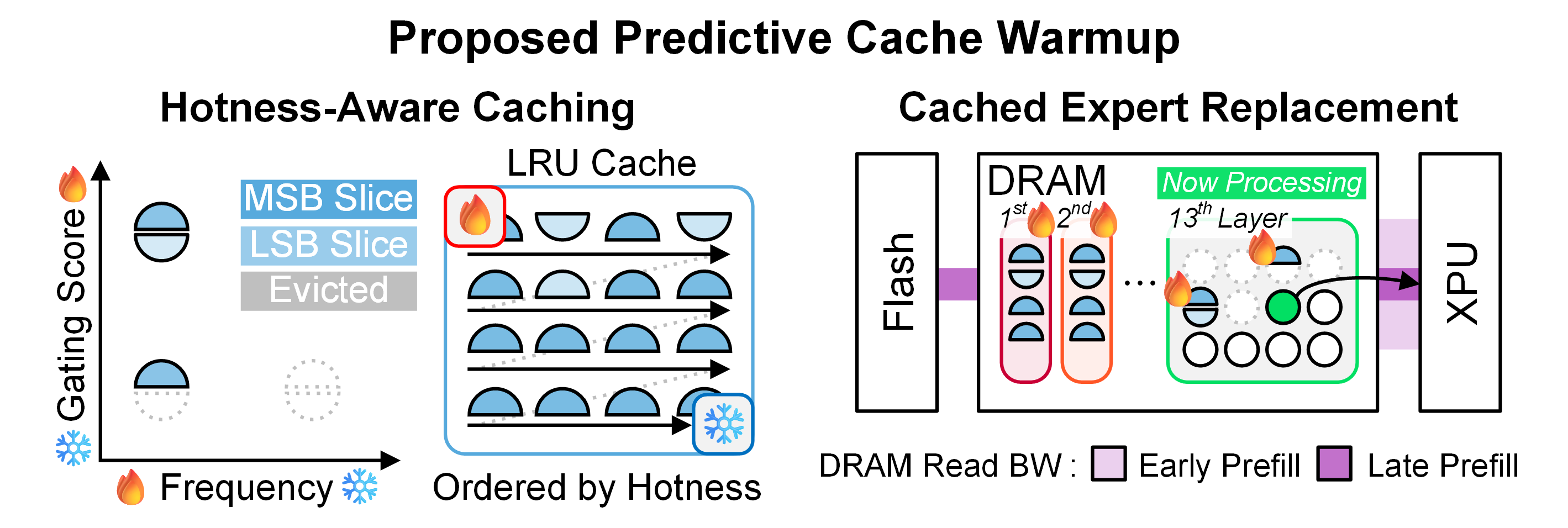}
        \caption{}
        \label{fig_5_b}
    \end{subfigure}

    \caption{(a) Caching challenges at the prefill-to-decode phase transition in single-batch execution. (b) Prefill-hotness–aware expert cache warmup strategy.}
    \label{fig_5}
    \vspace{-0.5cm}
\end{figure}

The core mechanism of cache warmup is to decide which slices to evict. In early prefill, DRAM bandwidth is sufficient, but as prefill progresses, pre-loaded experts have already been consumed and the system enters a regime where Flash bandwidth becomes comparable to XPU throughput. Prefetching ahead is no longer effective, as the system loads one expert from Flash while consuming another on the XPU. In this one-to-one exchange phase, it is essential to preserve the right slices from the currently processing layer.

Once prefill ends, the system re-orders the cache based on accumulated access frequency so that the resulting LRU state is aligned with experts expected to appear early in decode. As decode begins, this reordered cache is already primed with the right slices, substantially reducing cold misses, stabilizing performance, and enabling a smooth transition from prefill to decode without requiring retraining or complex scheduling—making the approach practical for real-world single-batch MoE deployments.

\section{System Implementation}

In our on-premises, single-batch MoE deployment scenario, the system consists of an XPU, main memory, and storage, as illustrated in Fig. 7. The XPU serves as the inference engine for model execution. Its systolic tensor PE array operates at 1 GHz and integrates 8,192 8-bit PEs, providing 16.4 TOPS of throughput with an energy efficiency of 3.18 TOPS/W. The main memory is LPDDR4 DRAM\cite{lpddr4}, offering approximately 104 Gbps of bandwidth and around 8 GB of capacity. Read/write operations consume up to 1.5 pJ/bit. In this configuration, DRAM is assumed to hold static model parameters. In practical mobile LLM deployments, however, DRAM is not dedicated exclusively to the model, requiring flexible and careful memory management. The storage layer is UFS 3.1 Flash\cite{ufs}, which provides a relatively low 10 Gbps bandwidth but a large 128 GB capacity, with an access energy of 103 pJ/bit. Flash stores the full set of model weights and is accessed only when an expert miss occurs. Due to its high energy cost and limited bandwidth, frequent Flash accesses can significantly degrade system-level performance.

\begin{figure}[t]
    \centering
    \includegraphics[width=\linewidth]{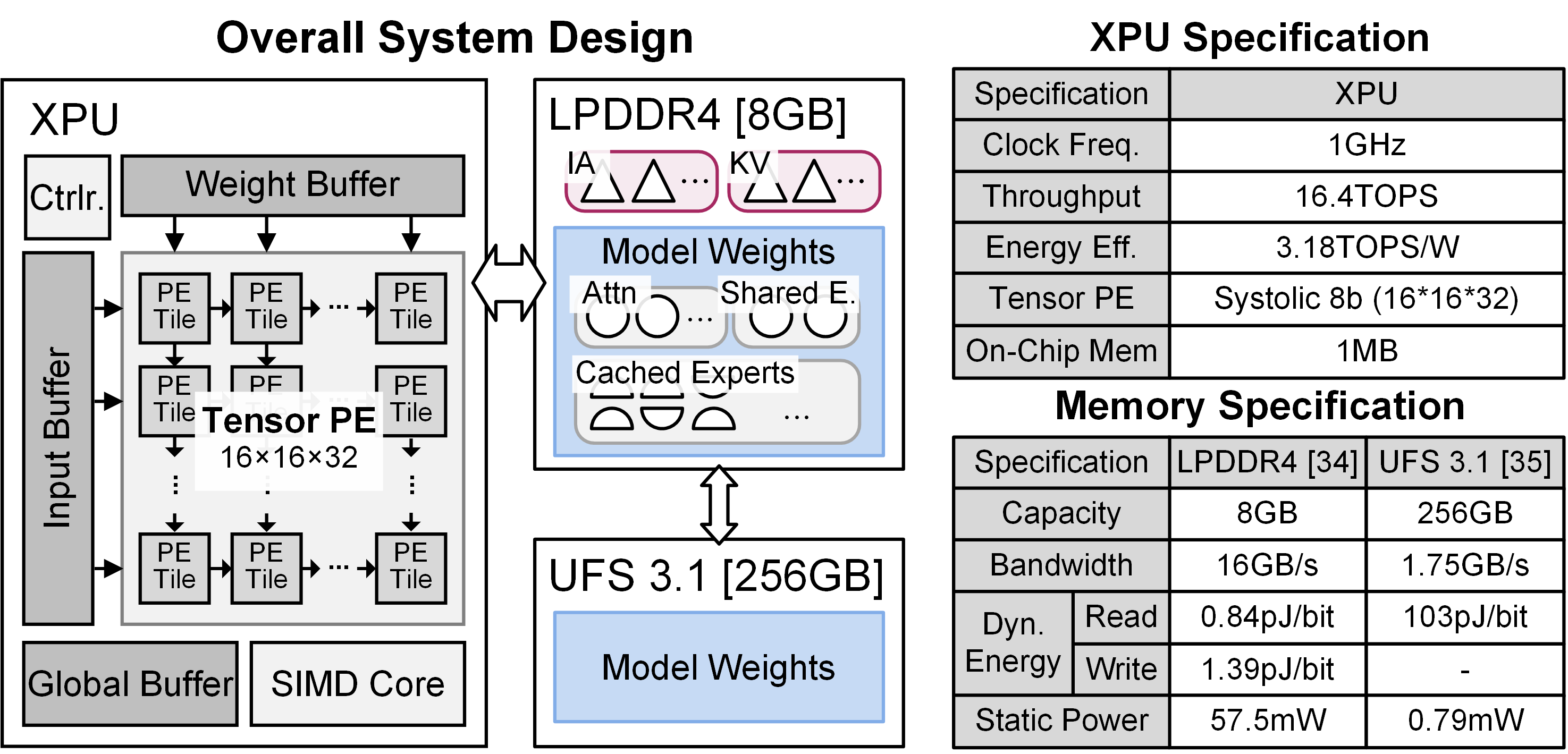}
    \caption{Overall SliceMoE System architecture and system specifications.}
    \label{fig_6}
    \vspace{-0.5cm}
\end{figure}

\section{Evaluation}

\subsection{Experimental Setup}

\textit{1) Models and Benchmark:} 
We evaluate SliceMoE on two MoE models with large expert pools: DeepSeek-V2-Lite~\cite{deepseek_v2} and Qwen1.5-MoE-A2.7B~\cite{qwen15_moe}. Since bit-sliced caching primarily affects the decode phase, we use GSM8K 5-shot~\cite{gsm8k}, which produces long outputs (prefill $\sim$500 tokens, decode $>$100 tokens). Accuracy is reported without prompt conditioning. We perform weight-only quantization, while keeping activations in FP16 and storing the KV cache in INT8. All non-expert model weights use G128 symmetric INT8 quantization, whereas expert weights use G32 asymmetric quantization, over which we sweep the precision configurations.

\textit{2) Baseline MoE Acceleration:}
SliceMoE is orthogonal to existing routing and caching methods. For routing, we compare against Cache-Prior—the state-of-the-art cache-aware routing method—and Cumsum routing, which is representative of cumulative threshold–based expert selection in high miss-rate regimes. This ensures evaluation across both locality-sensitive and locality-insensitive routing behaviors.

\begin{figure}[t]
    \centering
    \includegraphics[width=\linewidth]{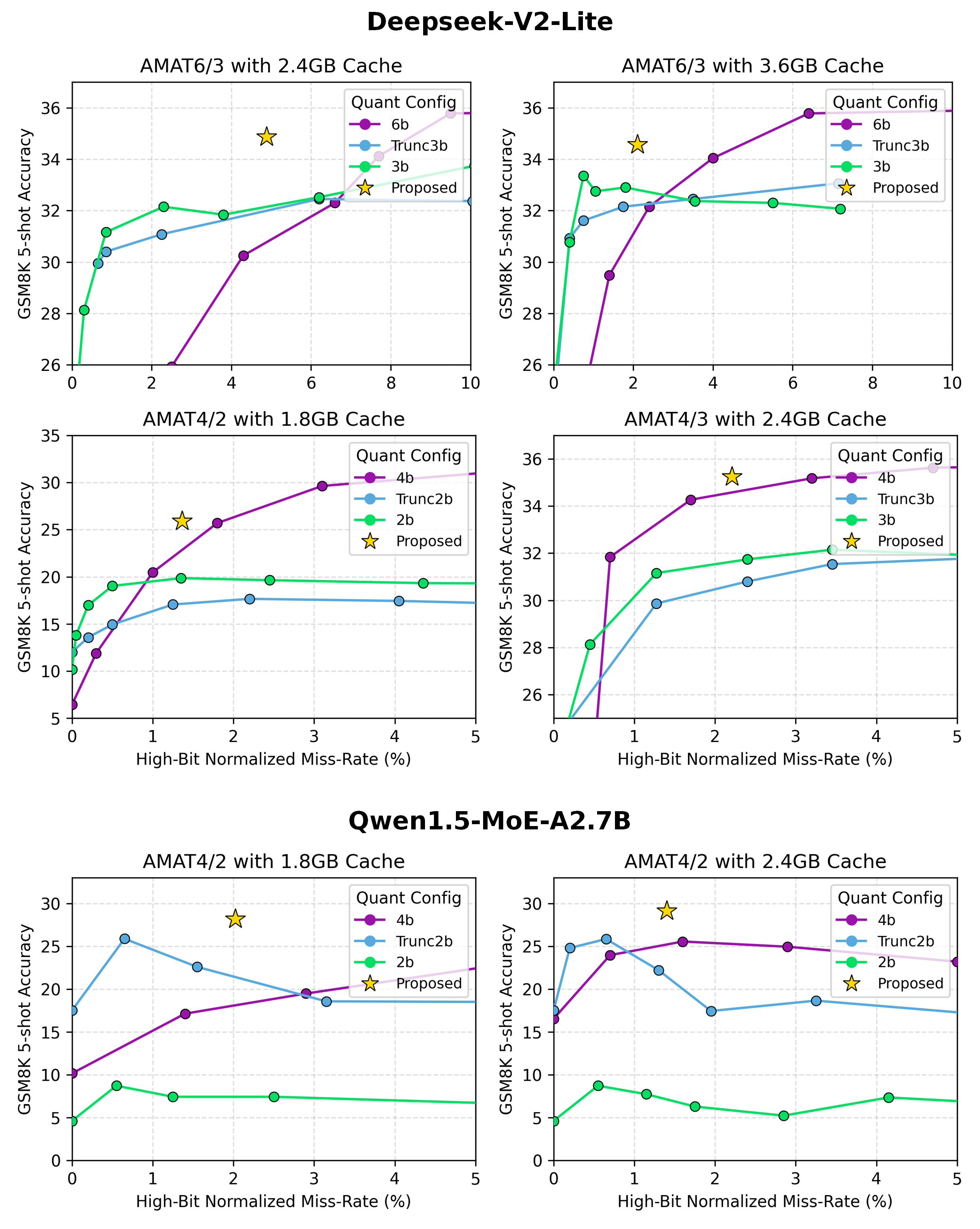}
    \caption{GSM8K 5-shot accuracy of DBSC under high-bit–normalized miss rates (no prompt conditioning).}
    \label{fig_7}
    \vspace{-0.5cm}
\end{figure}

\textit{3) Cache Implementation:}
We use a standard LRU cache to isolate the effect of bit-sliced caching from eviction heuristics. A unified cache is shared across layers because early layers exhibit wide expert usage while deeper layers show sharper distributions~\cite{laser}. To avoid initialization artifacts, the miss-rate constraint activates after the first 10 decode steps, providing a brief warm-up window.

\textit{4) Design Points:}
We consider three expert-only cache capacities—1.8 GB, 2.4 GB, and 3.6 GB—representing realistic on-device memory budgets. At 1.8 GB, at least one high-bit expert per layer fits; at 3.6 GB, fewer than half of all high-bit experts can be cached. Bitwidth configurations are chosen based on each model’s precision sensitivity; Qwen1.5-MoE-A2.7B is less sensitive than DeepSeek-V2-Lite, allowing slightly lower-bit experts while ensuring comparable accuracy.

\begin{figure*}[t]
    \centering
    \includegraphics[width=\linewidth]{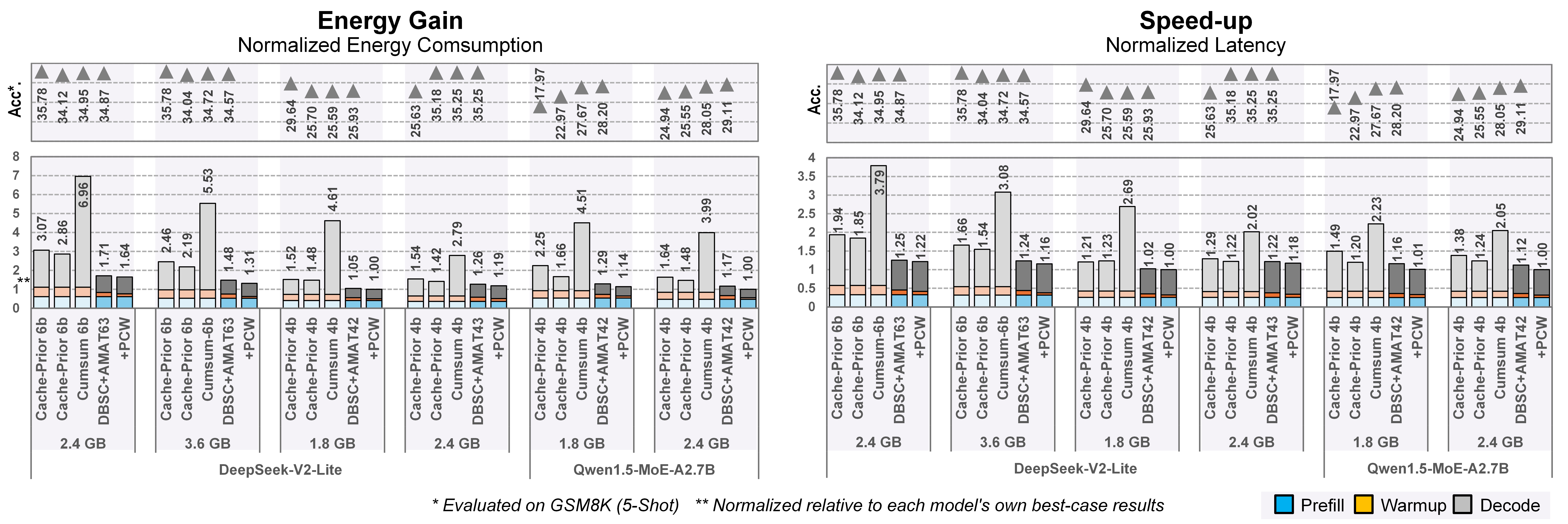}
    \caption{Energy gain and speed-up across cache-aware routing schemes on GSM8K with various cache configuration and bit-precision}
    \label{fig_9}
    \vspace{-0.5cm}
\end{figure*}

\subsection{Accuracy}

Fig. 8 summarizes GSM8K 5-shot accuracy as a function of the high-bit normalized miss rate under various miss-rate constraints. The visualization distinguishes several configurations: the high-bit baseline is shown in purple, the conventional low-bit configuration in green, and the AMAT-based mixed-precision setting—high-bit for prefill and low-bit for decode—in blue. The configuration combining AMAT with DBSC, indicated by the marker symbols, reflects the dynamic head–aware variant.

High-bit caching achieves strong accuracy at low miss rates, but its accuracy collapses once misses begin to accumulate due to its limited expert capacity. Conversely, aggressively low-bit caching maintains stable accuracy at extreme miss rates, yet it cannot surpass the inherent accuracy ceiling imposed by low precision—even when the cache is large and miss rate is nearly zero. AMAT-only configurations sit between these extremes—behaving like high-bit schemes at high miss rates and like low-bit schemes at low miss rates.

Our proposed DBSC + AMAT design consistently lies on the Pareto optimal. By reconstructing high-bit precision only for single-head–routed experts and using AMAT-derived low-bit slices for the rest, DBSC preserves substantial resolution while expanding the effective cache capacity. This avoids both major limitations: the sharp accuracy drop of high-bit caching under tight capacity and the inherent ceiling of uniform low-bit caching. As a result, the proposed configuration achieves higher accuracy at a given miss rate—and lower miss rate at a given accuracy—than all other design points.

\subsection{Energy-Gain and Speed-up}

Fig. 9 compares hardware cost under matched-accuracy conditions. Across all cache sizes and both models, DBSC—with AMAT and optional PCW—consistently outperforms the high-bit Cache-Prior baseline. All results were normalized to each model’s best case. Since prefill inherently requires high-bit computation, most savings occur in the decode phase, where DBSC applies AMAT-based bitwidth reduction. While Cumsum achieves strong accuracy, it is prohibitively expensive and never competitive. In contrast, DBSC delivers substantial efficiency gains, achieving up to 2.37× decode-stage energy reduction and 1.81× speed-up on DeepSeek-V2-Lite, and up to 2.85× energy reduction and 1.64× speed-up on Qwen1.5-MoE-A2.7B. These results show that DBSC occupies the most efficient point in the design space while preserving near–high-bit accuracy.

Fig. 10 further highlights the benefit of PCW by comparing cache initialization strategies. Warmup reshapes the cache into a hotness-aligned state, leading to 2.31× lower decode energy and 1.96× speed-up over an empty cache. Moreover, PCW improves accuracy by preserving frequently accessed MSB slices, yielding the best accuracy among all initial states. These results show that PCW complements DBSC+AMAT by reducing early cold misses and stabilizing performance.

\begin{figure}[t]
    \centering
    \includegraphics[width=\linewidth]{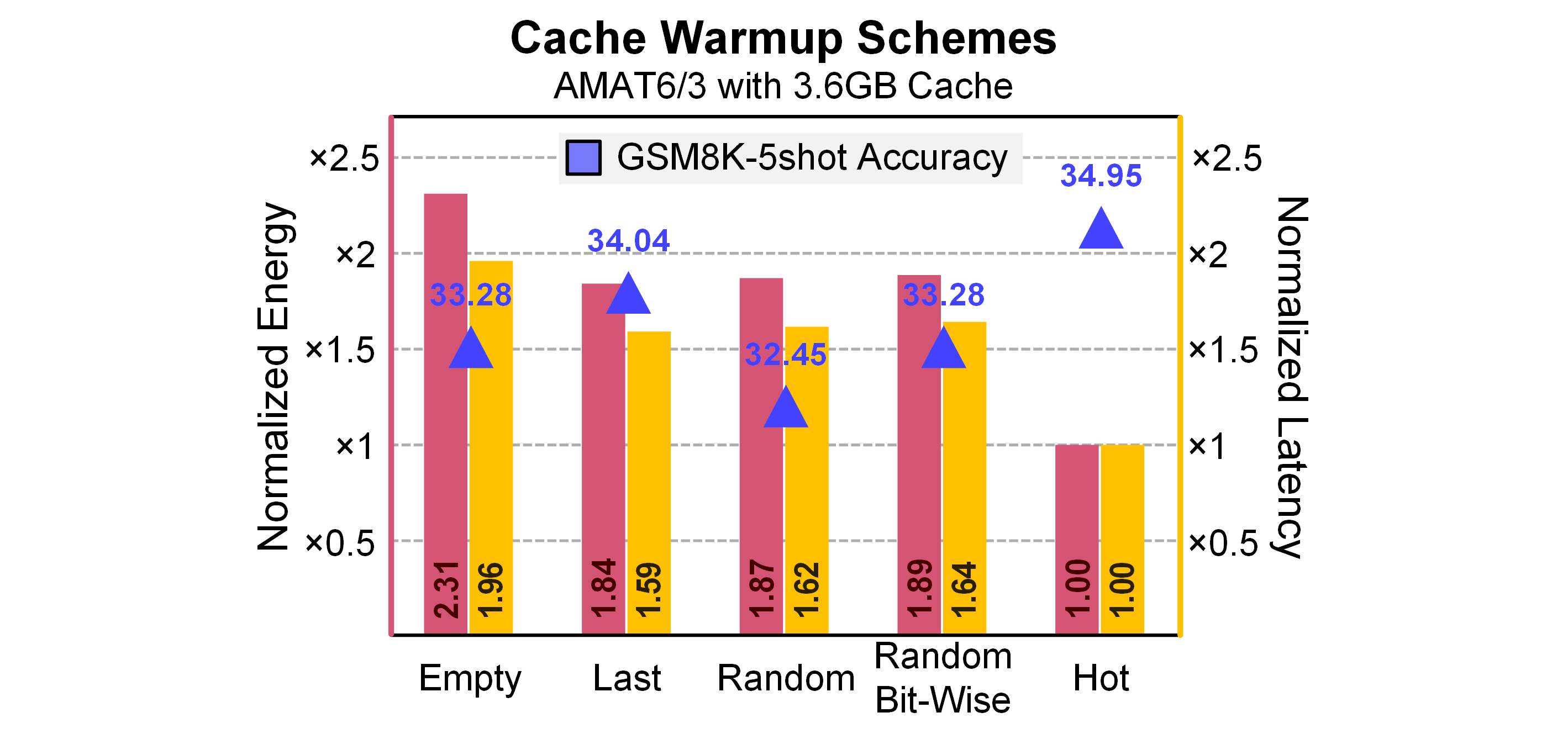}
    \caption{Cache warmup energy and latency across baseline cache initial state and PCW(hot)}
    \label{fig_8}
    \vspace{-0.5cm}
\end{figure}

\section{Conclusion}

In this work, we presented a unified framework for efficient single-batch MoE inference that combines Dynamic Bit-Sliced Caching (DBSC), Calibration-Free Asymmetric Matryoshka Quantization (AMAT), and Predictive Cache Warmup (PCW). DBSC adapts precision at slice granularity, significantly increasing cacheable expert capacity and improving hit behavior under tight memory budgets. AMAT provides a simple yet robust low-bit representation that remains compatible with high-bit slices, enabling substantial energy and latency reductions without sacrificing accuracy. PCW further bridges the gap between prefill and decode by reshaping the cache into a hotness-aligned state, reducing early cold misses and boosting both efficiency and model quality. Together, these components form a practical, calibration-free solution that consistently delivers near–high-bit accuracy while significantly improving efficiency across diverse MoE architectures. SliceMoE reduces decode-stage energy consumption by up to 2.37× on DeepSeek-V2-Lite and 2.85× on Qwen1.5-MoE-A2.7B, and improves decode latency by up to 1.81× and 1.64×, respectively. These results demonstrate that fine-grained precision control and slice-aware caching enable an effective design space for real-world MoE deployment under constrained hardware environments.


\vspace{12pt}

\end{document}